\documentclass[Physsubmission, Phys]{SciPost}
\def\MGvATNLO{{\sc MadGraph5\_aMC@NLO}}
\def\cascade{{\sc Cascade}}

\def\Pb{\ensuremath{\rm{b}}}
\def\PZ{\ensuremath{\rm{Z}}}

\def\Pqb{\ensuremath{\rm{b }}}
\def\Paqb{\ensuremath{\bar{\rm{b }}}}

\def\TeV{\ensuremath{{\rm{TeV }}}}

\def\ktz{\ensuremath{k_{\rm T,0}}}

\newcommand{\fourfl}{4FLVN}
\newcommand{\fivefl}{5FLVN}

\hypersetup{
    colorlinks,
    linkcolor={red!50!black},
    citecolor={blue!50!black},
    urlcolor={blue!80!black}
}

\usepackage[bitstream-charter]{mathdesign}
\DeclareSymbolFont{usualmathcal}{OMS}{cmsy}{m}{n}
\DeclareSymbolFontAlphabet{\mathcal}{usualmathcal}

\begin{document}

\begin{center}{\Large \textbf{
Transverse Momentum Dependent and collinear densities based on Parton Branching method\\
}}\end{center}

\begin{center}
S.~Taheri~Monfared and 
H.~Jung
\end{center}

\begin{center}
 Deutsches Elektronen-Synchrotron DESY, Germany
\\
* sara.taheri.monfared@desy.de
\end{center}

\begin{center}
\today
\end{center}


\definecolor{palegray}{gray}{0.95}
\begin{center}
\colorbox{palegray}{
  \begin{tabular}{rr}
  \begin{minipage}{0.1\textwidth}
    \includegraphics[width=30mm]{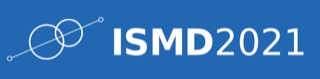}
  \end{minipage}
  &
  \begin{minipage}{0.75\textwidth}
    \begin{center}
    {\it 50th International Symposium on Multiparticle Dynamics}\\ {\it (ISMD2021)}\\
    {\it 12-16 July 2021} \\
    \doi{10.21468/SciPostPhysProc.?}\\
    \end{center}
  \end{minipage}
\end{tabular}
}
\end{center}

\section*{Abstract}
{\bf
We review the current status of collinear and Transverse Momentum Dependent densities based on the Parton Branching method. We investigate the performance of the PB-TMD evolution as well as PB-TMD parton shower with different configurations of the hard process in the four- and five-flavour schemes.}

\vspace{10pt}
\noindent\rule{\textwidth}{1pt}
\tableofcontents\thispagestyle{fancy}
\noindent\rule{\textwidth}{1pt}
\vspace{10pt}

\section{Introduction}
\label{sec:intro}
Perturbative Quantum Chromodynamics (QCD) Calculations have been performed at leading-order (LO), next-to-LO (NLO), and next-to-NLO (NNLO) accuracy in the strong coupling, and even to one order higher in some specific cases.  Since the work involved in these calculations increases sizably with the order, many more higher orders are not planned for near future.  An approximate solution is the parton shower in QCD Monte Carlo (MC) event generators using the soft and collinear approximation of partonic emissions. 
These algorithms are based on the law of large numbers and the central limit theorem and make use of random numbers. 
 In this approximation, enhanced terms in the perturbative expansion can be taken into account at all orders, the so called resummation of soft-collinear emissions.
Depending on which logarithms are resummed, we have
leading log (LL), next-to-leading log (NLL), etc.  
approximation. This plays an important role e.g. for the \PZ\ 
boson transverse momentum spectrum in Drell-Yan production: 
the  prediction from fixed-order perturbation theory diverges 
at low transverse momenta (where soft gluons contribute significantly) 
and only with resummation it becomes finite.

The collinear factorization theorem is the basis for the MC event generators in particle physics. So the perturbative treatment of partonic dynamics at short distances can be combined with non-pertubative modes of the hadronization process at large distance. One of the most developed method in this field is the Parton Branching (PB) method \cite{Hautmann:2017fcj,Hautmann:2017xtx}.

The PB formalism is based on the unitarity picture,  in terms of resolvable and non-resolvable branchings, and takes into account the role of the soft-gluon radiation and transverse momentum recoils in the evolution equations.
This method was successfully applied to describe the data from deep inelastic scattering at HERA \cite{BermudezMartinez:2018fsv} and DY transverse momentum spectra at LHC energies \cite{BermudezMartinez:2019anj} and fixed-target energies \cite{BermudezMartinez:2020tys}. 
TMD parton densities can be extracted from fits to any data sets applying the PB method for the parton evolution. 
In Refs. \cite{BermudezMartinez:2018fsv,Jung:2021vym}, the initial parton distributions were determined from a fit to HERA I+II neutral current and charged current inclusive DIS cross-section measurements at 5FLVN and 4FLVN schemes.

\section{PB-TMD parton densities}
Parton distributions are fundamental
tools to interpret experimental data for different hard-scattering processes considering underlying theory. Such processes are measured with the greatest precision by different
experiments around the world. In many
experiments, the precision of the measurement is higher than of the theory. A careful determination of PDFs and their uncertainties is mandatory to increase the precision of predictions. One of the limitations of the mainstream approach is the neglect of transverse degrees of freedom in a proton.

\subsection{PB-TMD parton densities in 5FLVN scheme}
Collinear and transverse momentum dependent (TMD) parton densities have been determined in the 5FLVN-scheme applying the PB method at exclusive level. As described in \cite{BermudezMartinez:2018fsv}, the initial parton distributions determined from a fit to inclusive deep inelastic scattering (DIS) cross section measurements at NLO with $m_b$=4.5~GeV and $\alpha_s(m_Z^{(n_f=5)})=0.118$.

The TMD parton densities are related to the collinear densities by 
\begin{equation}
f_{0,b} (x,{k_{t,0}^2},\mu_0^2) = f_{0,b} (x,\mu_0^2) \cdot \exp(-| \ktz^2 | /2 \sigma^2)  \;\;  
\end{equation}
where the intrinsic $\ktz^2$ distribution is given by a Gauss distribution with $\sigma^2  =  q_s^2 / 2$ at fixed 
$q_s = 0.5 ~ $GeV.  
Sensitivity to intrinsic, non-perturbative transverse momentum contributions were also checked with the fixed target DY data \cite{BermudezMartinez:2020tys}. The best width of the intrinsic $k_\perp$ gaussian is close to $0.5$ GeV, the value initially chosen in \cite{BermudezMartinez:2018fsv}. 

\subsection{PB-TMD parton densities in 4FLVN scheme}
The bottom quark does not appear as an active flavor in the 4FLVN scheme. We set $\alpha_s(m_Z^{(n_f=4)})$ $=0.1128$ in the evolution. The functional forms of the initial distribution are the same as what are used in the 5FLVN scheme, while the parameters of the PDFs are fitted to the same data set as used for 5FLVN scheme.

Comparing the 4FLVN and 5FLVN collinear PDFs, as shown in Fig. 2 and Fig. 3 of Ref. \cite{Jung:2021vym}, we observe 5FLVN PDFs are slightly smaller due to presence of more active flavors in the hard process. The same behaviour is reflected in TMD PDFs at small transverse momenta. At large $k_\perp$, the 4FLVN and 5FLVN TMD PDFs are identical.

\section{Showering in different schemes}
The outstanding benefit of the TMD PB method is that the evolution that it provides for TMD PDFs is performed explicitly. This feature is crucial for the consistent calculation of fully exclusive processes as it lets to construct parton showers which exactly follow the parton density evolution. So when the TMD distributions evaluated at the evolution scale is determined, the corresponding TMD parton shower can be precisely generated by backward evolution.
  
In 4FLVN scheme, bottom quark mass effects are retained in the computation of the hard cross section while associated PDFs do not contain any bottom distributions thus forbidding bottom-quark initiated contributions. The 5FLVN scheme instead neglects bottom-quark mass effects in the hard cross
section but allows the bottom PDFs to be radiatively produced by evolution. As a consequence, this scheme does include bottom-quark initiated contributions.

The process under consideration, $\PZ+ \Pqb\Paqb$ jets in proton-proton collisions, is simulated using \MGvATNLO\ ~\cite{Frixione:2006gn} interfaced to HERWIG6 in the 4- and 5-FLVN schemes using sets of (TMD) PDFs obtained through the PB approach in the respective
two schemes. 
Having determined the collinear and TMD PDFs via xFitter \cite{Alekhin:2014irh}, we studied the contribution from the TMD evolution to $\PZ+ \Pqb\Paqb$ jet production using the backward formulation of the PB TMD evolution equation implemented in the \cascade\ event generator \cite{Baranov:2021uol}.

\begin{figure}

\begin{minipage}{0.5\linewidth}
\centerline{\includegraphics[width=0.9\linewidth]{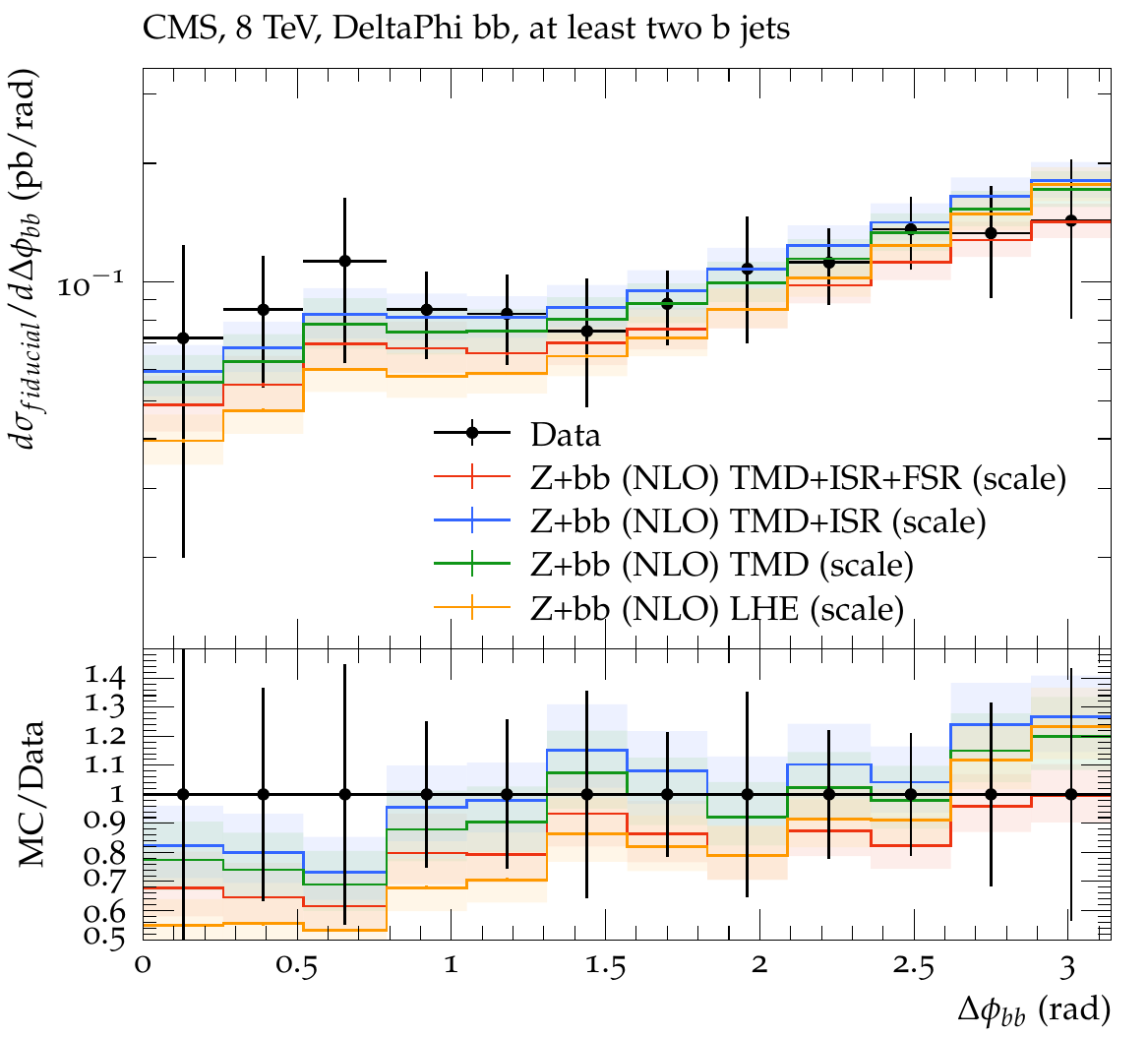}}
\caption*{(a) 4FLVN}
\end{minipage}
\hfill
\begin{minipage}{0.5\linewidth}
\centerline{\includegraphics[width=0.9\linewidth]{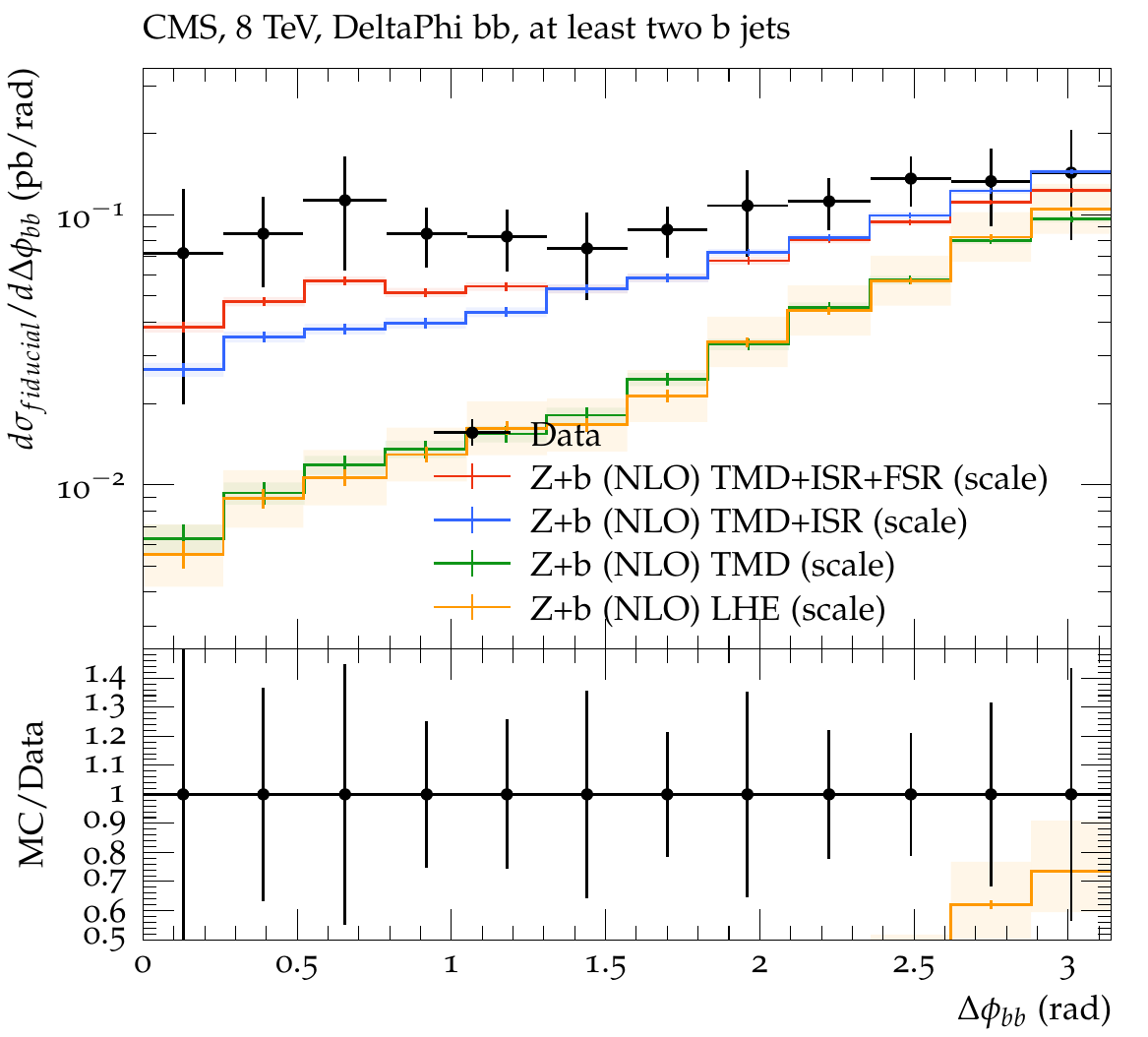}}
\caption*{(b) 5FLVN}
\end{minipage}
\caption{Differential cross section for $\PZ+ \Pqb\Paqb$ tagged jets as a function of azimuthal angular separation as measured by CMS \protect\cite{CMS:2016gmz} at $\sqrt{s}=8$ \TeV. The \fourfl -prediction is shown in (a), the \fivefl -prediction in (b).  The LHE files (parton level), after inclusion of PB-TMDs, initial state parton shower and final state parton shower are illustrated separately.
 }
\label{fig:ptb-twob}
\end{figure}

\begin{figure}
\begin{minipage}{0.5\linewidth}
\centerline{\includegraphics[width=0.9\linewidth]{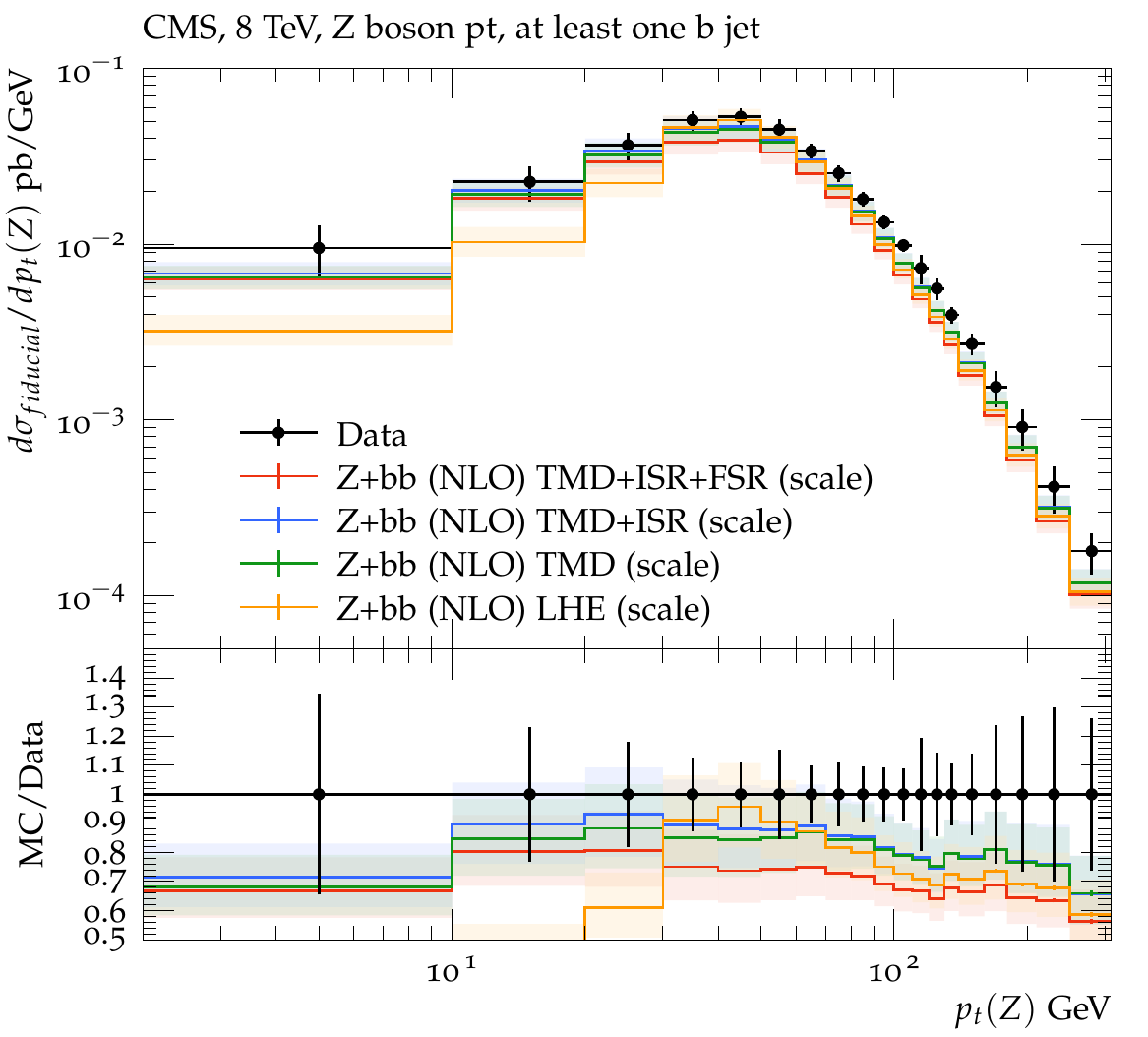}}
\caption*{(a) 4FLVN}
\end{minipage}
\hfill
\begin{minipage}{0.5\linewidth}
\centerline{\includegraphics[width=0.9\linewidth]{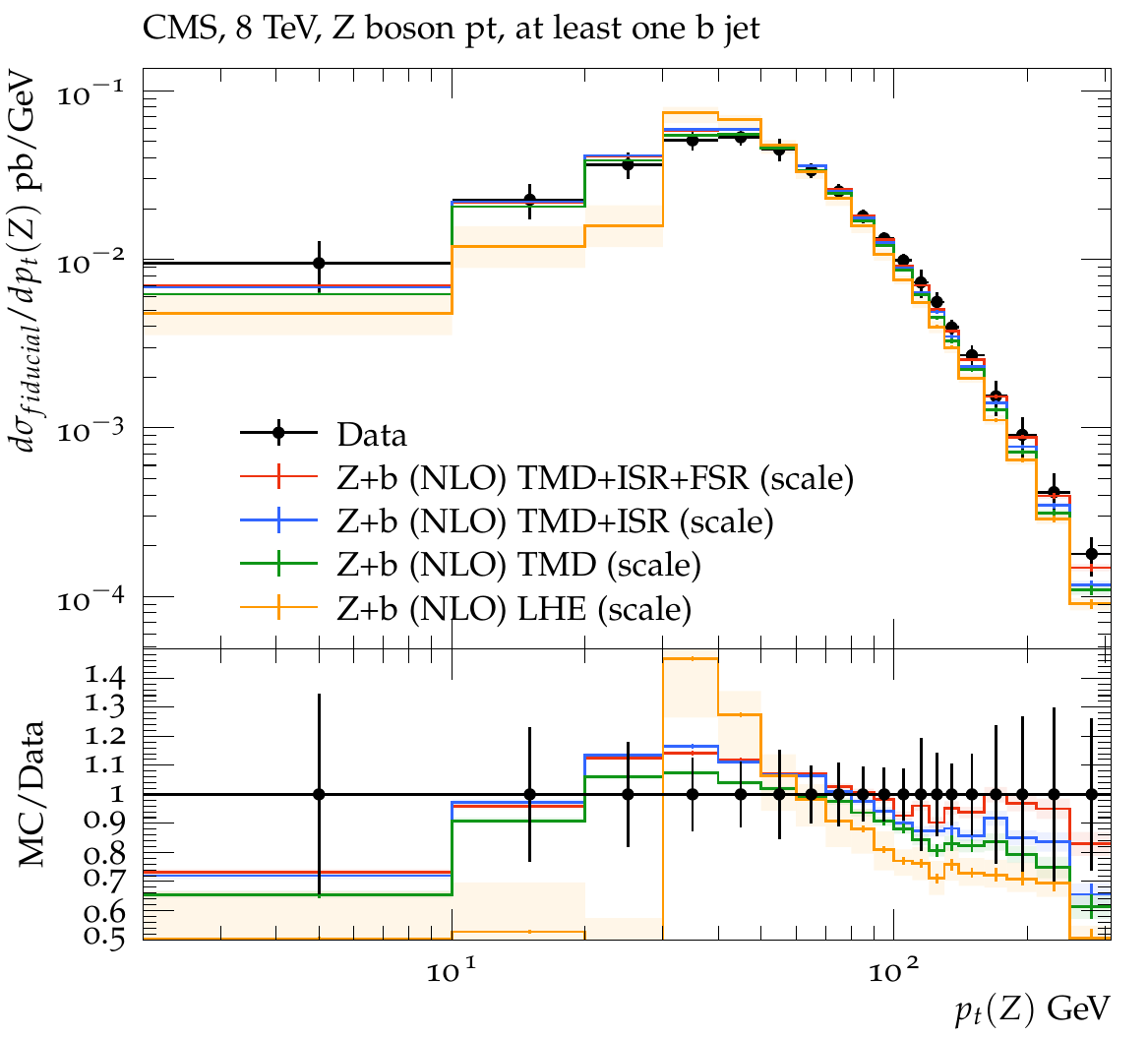}}
\caption*{(b) 5FLVN}
\end{minipage}
\caption[]{Differential cross section for $\PZ+ \Pqb\Paqb$ tagged jets as a function of the transverse momentum of the \PZ -boson as measured by CMS \protect\cite{CMS:2016gmz} at $\sqrt{s}=8$ \TeV. The \fourfl -prediction is shown in (a), the \fivefl -prediction in (b).  The LHE files (parton level), after inclusion of PB-TMDs, initial state parton shower and final state parton shower are illustrated separately.}
\label{fig:ptz-twob}
\end{figure}

The observables considered are the azimuthal angular separation between the two \Pb -tagged jets in $\PZ+ \Pqb\Paqb$ events  in Fig.~\ref{fig:ptb-twob} and the transverse momentum of \PZ -boson in Fig.~\ref{fig:ptz-twob} 
as measured by CMS \protect\cite{CMS:2016gmz} at $\sqrt{s}=8$ \TeV.
In yellow, the matrix element calculation is shown and in green the same calculation includes the TMD contribution. In addition, the blue curve includes initial-state radiation, while in red the final
result including final-state radiation as provided by the PYTHIA6 generator \cite{Sjostrand:2006za}. Gluon fusion to $\PZ+ \Pqb\Paqb$ is considered in final-state radiation in both schemes.
 The contribution
from initial-state radiation provided by the TMD evolution results is crucial for the description of transverse momentum of \PZ -boson and \Pb -quark in 5FLVN scheme, while 4FLVN calculation only weakly depends on PB-TMD and parton shower. Since both b partons are already produced with NLO accuracy at the matrix element level.

The TMD approach, as opposed to the purely collinear one, is more accurate in specific regions of the phase space, typically when hard radiation is inhibited. In both schemes in Fig.~\ref{fig:ptb-twob} and Fig.~\ref{fig:ptz-twob}, final results can well-describe the measurements at large $\Delta_{\phi}(b\bar{b})$ and small $p_t(Z)$ regions.

\section{Conclusion}
We studied $\PZ+ \Pqb\Paqb$ tagged jets production using
TMD densities and TMD showers produced in two different schemes of 4FLVN and 5FLNS with the parton branching approach. 
Both predictions are in very good agreement with measurements obtained by CMS at $\sqrt{s}=8$ \TeV. 

\section*{Acknowledgements}
We thank A.~Bermudez Martinez, L.I.~ Estevez Banos,	F.~Hautmann, M.~M.~Morentin, A.~Lelek,  Q.~Wang, H.~Yang for collaboration and discussion.
 STM thanks the Humboldt Foundation for the Georg Forster research fellowship.






\providecommand{\href}[2]{#2}\begingroup\raggedright\begin{thebibliography}{10}

\bibitem{Angeles-Martinez:2015sea}
R.~Angeles-Martinez et~al., \emph{{Transverse Momentum Dependent (TMD) parton
  distribution functions: status and prospects}},
  \href{https://doi.org/10.5506/APhysPolB.46.2501}{\emph{Acta Phys. Polon. B}
  {\bfseries 46} (2015) 2501}
  [\href{https://arxiv.org/abs/1507.05267}{{\ttfamily 1507.05267}}].

\bibitem{HAUTMANN2017446}
F.~Hautmann, H.~Jung, A.~Lelek, V.~Radescu and R.~Žlebčík, \emph{Soft-gluon
  resolution scale in qcd evolution equations},
  \href{https://doi.org/https://doi.org/10.1016/j.physletb.2017.07.005}{\emph{Physics
  Letters B} {\bfseries 772} (2017) 446}.

\bibitem{Hautmann2018}
F.~Hautmann, H.~Jung, A.~Lelek, V.~Radescu and R.~{\v{Z}}leb{\v{c}}{\'i}k,
  \emph{Collinear and tmd quark and gluon densities from parton branching
  solution of qcd evolution equations},
  \href{https://doi.org/10.1007/JHEP01(2018)070}{\emph{Journal of High Energy
  Physics} {\bfseries 2018} (2018) 70}.

\bibitem{PhysRevD.99.074008}
A.B.~Martinez, P.~Connor, H.~Jung, A.~Lelek, R.~\ifmmode \check{Z}\else
  \v{Z}\fi{}leb\ifmmode~\check{c}\else \v{c}\fi{}\'{\i}k, F.~Hautmann et~al.,
  \emph{Collinear and tmd parton densities from fits to precision dis
  measurements in the parton branching method},
  \href{https://doi.org/10.1103/PhysRevD.99.074008}{\emph{Phys. Rev. D}
  {\bfseries 99} (2019) 074008}.

\bibitem{Alekhin2015}
S.~Alekhin, O.~Behnke, P.~Belov, S.~Borroni, M.~Botje, D.~Britzger et~al.,
  \emph{Herafitter},
  \href{https://doi.org/10.1140/epjc/s10052-015-3480-z}{\emph{The European
  Physical Journal C} {\bfseries 75} (2015) 304}.

\bibitem{Hautmann_2014}
F.~Hautmann, H.~Jung, M.~Krämer, P.J.~Mulders, E.R.~Nocera, T.C.~Rogers
  et~al., \emph{Tmdlib and tmdplotter: library and plotting tools for
  transverse-momentum-dependent parton distributions},
  \href{https://doi.org/10.1140/epjc/s10052-014-3220-9}{\emph{The European
  Physical Journal C} {\bfseries 74} (2014) }.

\bibitem{PhysRevD.100.074027}
A.~Bermudez~Martinez, P.L.S.~Connor, D.~Dominguez~Damiani, L.I.~Estevez~Banos,
  F.~Hautmann, H.~Jung et~al., \emph{Production of $z$ bosons in the parton
  branching method},
  \href{https://doi.org/10.1103/PhysRevD.100.074027}{\emph{Phys. Rev. D}
  {\bfseries 100} (2019) 074027}.

\bibitem{HAUTMANN2019114795}
F.~Hautmann, L.~Keersmaekers, A.~Lelek and A.~{van Kampen}, \emph{Dynamical
  resolution scale in transverse momentum distributions at the lhc},
  \href{https://doi.org/https://doi.org/10.1016/j.nuclphysb.2019.114795}{\emph{Nuclear
  Physics B} {\bfseries 949} (2019) 114795}.

\bibitem{Martinez2020}
A.B.~Martinez, P.L.S.~Connor, D.D.~Damiani, L.I.E.~Banos, F.~Hautmann, H.~Jung
  et~al., \emph{The transverse momentum spectrum of low mass drell--yan
  production at next-to-leading order in the parton branching method},
  \href{https://doi.org/10.1140/epjc/s10052-020-8136-y}{\emph{The European
  Physical Journal C} {\bfseries 80} (2020) 598}.

\bibitem{Abdulov_2021}
N.A.~Abdulov, A.~Bacchetta, S.~Baranov, A.~Bermudez~Martinez, V.~Bertone,
  C.~Bissolotti et~al., \emph{Tmdlib2 and tmdplotter: a platform for 3d hadron
  structure studies},
  \href{https://doi.org/10.1140/epjc/s10052-021-09508-8}{\emph{The European
  Physical Journal C} {\bfseries 81} (2021) }.

\bibitem{Jung2010}
H.~Jung, S.~Baranov, M.~Deak, A.~Grebenyuk, F.~Hautmann, M.~Hentschinski
  et~al., \emph{The ccfm monte carlo generator cascade version 2.2.03},
  \href{https://doi.org/10.1140/epjc/s10052-010-1507-z}{\emph{The European
  Physical Journal C} {\bfseries 70} (2010) 1237}.

\bibitem{Baranov2021}
S.~Baranov, A.~Bermudez~Martinez, L.I.~Estevez~Banos, F.~Guzman, F.~Hautmann,
  H.~Jung et~al., \emph{Cascade3 a monte carlo event generator based on tmds},
  \href{https://doi.org/10.1140/epjc/s10052-021-09203-8}{\emph{The European
  Physical Journal C} {\bfseries 81} (2021) 425}.


\bibitem{Martinez:2021chk}
A.B.~Martinez, F.~Hautmann and M.L.~Mangano, \emph{{TMD Evolution and Multi-Jet
  Merging}},  \href{https://arxiv.org/abs/2107.01224}{{\ttfamily 2107.01224}}.


\bibitem{Jung:2021mox}
H.~Jung, S.~Taheri Monfared and T.~Wening,
\emph{Determination of collinear and TMD photon densities using the Parton Branching method}, 
Phys. Lett. B \textbf{817} (2021), 136299
doi:10.1016/j.physletb.2021.136299
[arXiv:2102.01494 [hep-ph]].


\bibitem{CMS:2018mdl}
A.~M.~Sirunyan \textit{et al.} [CMS],
\emph{Measurement of the differential Drell-Yan cross section in proton-proton collisions at $ \sqrt{\mathrm{s}} $ = 13 TeV}, 
JHEP \textbf{12} (2019), 059
doi:10.1007/JHEP12(2019)059
[arXiv:1812.10529 [hep-ex]].



\bibitem{Jung:2021vym}
H.~Jung and S.~T.~Monfared,
\emph{TMD parton densities and corresponding parton showers: the advantage of four- and five-flavour schemes},
[arXiv:2106.09791 [hep-ph]].

\bibitem{BermudezMartinez:2018fsv}
A.~Bermudez Martinez, P.~Connor, H.~Jung, A.~Lelek, R.~\v{Z}leb\v{c}\'\i{}k, F.~Hautmann and V.~Radescu,
\emph{Collinear and TMD parton densities from fits to precision DIS measurements in the parton branching method}, 
Phys. Rev. D \textbf{99} (2019) no.7, 074008
doi:10.1103/PhysRevD.99.074008
[arXiv:1804.11152 [hep-ph]].


\bibitem{CATANI1994475}
S.~Catani and F.~Hautmann, \emph{High-energy factorization and small-x deep
  inelastic scattering beyond leading order},
  \href{https://doi.org/https://doi.org/10.1016/0550-3213(94)90636-X}{\emph{Nuclear
  Physics B} {\bfseries 427} (1994) 475}.

\bibitem{Hautmann:2012sh}
F.~Hautmann, M.~Hentschinski and H.~Jung, \emph{{Forward Z-boson production and
  the unintegrated sea quark density}},
  \href{https://doi.org/10.1016/j.nuclphysb.2012.07.023}{\emph{Nucl. Phys. B}
  {\bfseries 865} (2012) 54} [\href{https://arxiv.org/abs/1205.1759}{{\ttfamily
  1205.1759}}].

\bibitem{Gituliar2016}
O.~Gituliar, M.~Hentschinski and K.~Kutak, \emph{Transverse-momentum-dependent
  quark splitting functions in kt-factorization: real contributions},
  \href{https://doi.org/10.1007/JHEP01(2016)181}{\emph{Journal of High Energy
  Physics} {\bfseries 2016} (2016) 181}.

\bibitem{Hentschinski:2016wya}
M.~Hentschinski, A.~Kusina and K.~Kutak, \emph{{Transverse momentum dependent
  splitting functions at work: quark-to-gluon splitting}},
  \href{https://doi.org/10.1103/PhysRevD.94.114013}{\emph{Phys. Rev. D}
  {\bfseries 94} (2016) 114013}
  [\href{https://arxiv.org/abs/1607.01507}{{\ttfamily 1607.01507}}].

\bibitem{Hentschinski2018}
M.~Hentschinski, A.~Kusina, K.~Kutak and M.~Serino, \emph{Tmd splitting
  functions in $k_t$ factorization: the real contribution to the gluon-to-gluon
  splitting}, \href{https://doi.org/10.1140/epjc/s10052-018-5634-2}{\emph{The
  European Physical Journal C} {\bfseries 78} (2018) 174}.

\bibitem{Hentschinski:2021lsh}
M.~Hentschinski, \emph{{Transverse Momentum Dependent Gluon Distribution within
  High Energy Factorization at Next-to-Leading Order}},
  \href{https://arxiv.org/abs/2107.06203}{{\ttfamily 2107.06203}}.

\bibitem{keersmaekers2021implementing}
L.~Keersmaekers, \emph{Implementing transverse momentum dependent splitting
  functions in parton branching evolution equations},
  \href{https://arxiv.org/abs/2109.07326}{{\ttfamily 2109.07326}}.

\bibitem{lissaetal-inprep}
L.~Keersmaekers et~al. In preparation.

\end{thebibliography}\endgroup


\begin{thebibliography}{99}
\bibitem{Hautmann:2017fcj}
F.~Hautmann, H.~Jung, A.~Lelek, V.~Radescu and R.~Zlebcik,
JHEP \textbf{01} (2018), 070
doi:10.1007/JHEP01(2018)070
[arXiv:1708.03279 [hep-ph]].

\bibitem{Hautmann:2017xtx}
F.~Hautmann, H.~Jung, A.~Lelek, V.~Radescu and R.~Zlebcik,
Phys. Lett. B \textbf{772} (2017), 446-451
doi:10.1016/j.physletb.2017.07.005
[arXiv:1704.01757 [hep-ph]].





\bibitem{BermudezMartinez:2018fsv}
A.~Bermudez Martinez, P.~Connor, H.~Jung, A.~Lelek, R.~\v{Z}leb\v{c}\'\i{}k, F.~Hautmann and V.~Radescu,
Phys. Rev. D \textbf{99} (2019) no.7, 074008
doi:10.1103/PhysRevD.99.074008
[arXiv:1804.11152 [hep-ph]].


\bibitem{BermudezMartinez:2019anj}
A.~Bermudez Martinez, P.~Connor, D.~Dominguez Damiani, L.~I.~Estevez Banos, F.~Hautmann, H.~Jung, J.~Lidrych, M.~Schmitz, S.~Taheri Monfared and Q.~Wang, \textit{et al.}
Phys. Rev. D \textbf{100} (2019) no.7, 074027
doi:10.1103/PhysRevD.100.074027
[arXiv:1906.00919 [hep-ph]].


\bibitem{BermudezMartinez:2020tys}
A.~Bermudez Martinez, P.~L.~S.~Connor, D.~Dominguez Damiani, L.~I.~Estevez Banos, F.~Hautmann, H.~Jung, J.~Lidrych, A.~Lelek, M.~Mendizabal and M.~Schmitz, \textit{et al.}
Eur. Phys. J. C \textbf{80} (2020) no.7, 598
doi:10.1140/epjc/s10052-020-8136-y
[arXiv:2001.06488 [hep-ph]].

\bibitem{Jung:2021vym}
H.~Jung and S.~T.~Monfared,
[arXiv:2106.09791 [hep-ph]].

\bibitem{Frixione:2006gn}
S.~Frixione and B.~R.~Webber,
[arXiv:hep-ph/0612272 [hep-ph]].

\bibitem{Alekhin:2014irh}
S.~Alekhin, O.~Behnke, P.~Belov, S.~Borroni, M.~Botje, D.~Britzger, S.~Camarda, A.~M.~Cooper-Sarkar, K.~Daum and C.~Diaconu, \textit{et al.}
Eur. Phys. J. C \textbf{75} (2015) no.7, 304
doi:10.1140/epjc/s10052-015-3480-z
[arXiv:1410.4412 [hep-ph]].


\bibitem{Baranov:2021uol}
S.~Baranov, A.~Bermudez Martinez, L.~I.~Estevez Banos, F.~Guzman, F.~Hautmann, H.~Jung, A.~Lelek, J.~Lidrych, A.~Lipatov and M.~Malyshev, \textit{et al.}
Eur. Phys. J. C \textbf{81} (2021) no.5, 425
doi:10.1140/epjc/s10052-021-09203-8
[arXiv:2101.10221 [hep-ph]].


\bibitem{CMS:2016gmz}
V.~Khachatryan \textit{et al.} [CMS],
Eur. Phys. J. C \textbf{77} (2017) no.11, 751
doi:10.1140/epjc/s10052-017-5140-y
[arXiv:1611.06507 [hep-ex]].

\bibitem{Sjostrand:2006za}
T.~Sjostrand, S.~Mrenna and P.~Z.~Skands,
JHEP \textbf{05} (2006), 026
doi:10.1088/1126-6708/2006/05/026
[arXiv:hep-ph/0603175 [hep-ph]].
\end{thebibliography}

\nolinenumbers

\end{document}